# Personnalisation de Systèmes OLAP Annotés

H. Jerbi, G. Pujolle, F. Ravat, O. Teste

*Université de Toulouse, IRIT (UMR5505)*
*118, Route de Narbonne - F-31062 Toulouse cedex 9*
*{jerbi, pujolle, ravat, teste}@irit.fr*

*RÉSUMÉ. La problématique traitée dans cet article consiste à personnaliser les systèmes OLAP annotés. Nous proposons de modéliser les données au sein d'une constellation supportant à la fois des annotations et des préférences. Les annotations sont utilisées pour représenter « l'expertise » immatérielle du décideur tandis que les préférences permettent d'individualiser les données durant les manipulations OLAP. Les préférences sont exploitées pour des recommandations contextuelles annotées qui assistent l'usager dans son exploration de l'espace multidimensionnel.*

*ABSTRACT. This paper deals with personalization of annotated OLAP systems. Data constellation is extended to support annotations and user preferences. Annotations reflect the decision-maker experience whereas user preferences enable users to focus on the most interesting data. User preferences allow annotated contextual recommendations helping the decision-maker during his/her multidimensional navigations.*

MOTS-CLÉS : *Système d'aide à la décision, Bases de données multidimensionnelles, Personnalisation.*

## 1. Introduction

Les systèmes OLAP (« On-Line Analytical Processing ») facilitent l'analyse en offrant un espace de représentation multidimensionnelle des données que les décideurs explorent interactivement. Cette approche a connu un développement important grâce à sa capacité à permettre un accès direct et interactif aux données analysées. Cependant ces systèmes sont élaborés pour un groupe de décideurs ou un sujet d'analyse (« subject-oriented » (Inmon, 1994)) pour lesquels sont présumés des besoins parfaitement identiques. Cette simplification rend les systèmes OLAP parfois mal adaptés à un usage particulier. Le décideur se trouve confronté à un espace multidimensionnel, souvent très vaste, sur lequel il doit opérer un nombre important de manipulations afin d'obtenir un résultat le plus proche possible de son besoin.

La problématique traitée dans cet article consiste à personnaliser les systèmes OLAP en fonction de besoins analytiques individuels. Les mécanismes de personnalisation dans les systèmes OLAP, visant à mieux prendre en compte l'usager, ne sont que très peu étudiés. Nous proposons d'utiliser la personnalisation afin de rendre possible au décideur des recommandations de manipulations OLAP. L'originalité de l'approche proposée réside dans le couplage des recommandations avec un processus d'annotation qui permet de prendre en compte non seulement des données brutes manipulées mais également des réflexions, des interprétations et des commentaires des décideurs. Les systèmes OLAP sont généralement réduits à la mise à disposition des données décisionnelles tandis que les décideurs doivent analyser ces données en s'appuyant sur la base immatérielle de leur expertise.

Après analyse des différents travaux relatifs à la personnalisation des systèmes OLAP (section 2), nous présentons notre modèle OLAP (section 3), les annotations (section 4), les concepts de personnalisation retenus (section 5) et l'implantation combinée des mécanismes de personnalisation et d'annotations (section 6).

## 2. Etat de l'art

La personnalisation d'un système consiste à définir, puis à exploiter un profil utilisateur (Korfhage, 1997) pouvant s'apparenter à une modélisation de l'usager/décideur. Aucun consensus n'existe sur la définition de profil ; on peut relever cependant la proposition d'un profil générique multidimensionnel visant à couvrir une majorité de contextes (Bouzeghoub *et al.*, 2005). Nous définissons un profil comme un ensemble de caractéristiques servant à configurer ou à adapter le système à l'utilisateur, afin de lui fournir des réponses plus adaptées afin de faciliter sa tâche d'analyse de données. La définition explicite d'un profil (l'utilisateur doit effectuer des interactions avec le système) correspond à la configuration (paramétrage) d'un système tandis que la définition implicite (le système s'adapte automatiquement à l'utilisateur) s'apparente à l'apprentissage. La configuration consiste pour l'utilisateur à paramétrer explicitement son profil tandis que l'adaptation consiste pour le système à définir implicitement le profil de l'utilisateur.

Alors que la personnalisation a fait l'objet de très nombreux travaux en recherche d'information et bases de données (Ioannidis *et al.*, 2005), très peu de propositions visent à personnaliser les systèmes OLAP (Rizzi, 2007) (Bentayeb *et al.*, 2009). Plusieurs travaux (Sapia, 2000) (Talhammer *et al.*, 2001) (Bellatreche *et al.*, 2009) (Favre *et al.*, 2007) visent à transformer le système OLAP en fonction de préférences explicitement collectées (Golfarelli *et al.*, 2009). Seuls les travaux dans (Giacometti *et al.*, 2008, 2009) consistent à recommander des requêtes sur la base de calculs opérés sur l'historique des navigations réalisées par un groupe d'utilisateurs. Ces travaux traitent des requêtes anticipées et recommandent des requêtes complètes déjà jouées par le groupe social d'usagers, introduisant des approximations pouvant être importantes. Récemment, les auteurs dans (Garrigos *et al.*, 2009) présentent une approche pour personnaliser les bases de données multidimensionnelles à un niveau d'abstraction conceptuel comme nous le proposons. Ces travaux modélisent le décideur de manière exhaustive (définition d'un profil UML) et la personnalisation est basée sur l'utilisation d'évènements (fonctions représentant les opérations OLAP telles que forage, sélection, rotation…) et de conditions (fonctions mettant à jour les données en fonction des évènements et du profil utilisateur). Ces travaux se focalisent sur la personnalisation des données brutes entreposées dans les systèmes OLAP sans prendre en compte l'expertise des décideurs. De plus, ces travaux ne permettent pas de "contextualiser" les préférences utilisateurs comme nous le proposons.

**3. De la constellation à son analyse**

Nous proposons de modéliser les données dans un système OLAP par extension du concept de constellation (Ravat *et al.*, 2007) afin de permettre sa personnalisation et d'associer des annotations aux composants de celle-ci. Cette section explicite les concepts OLAP de modélisation des données et les principes retenus pour modéliser une analyse multidimensionnelle.

**3.1. *Schéma en constellation***

Soient $F$ l'ensemble des faits, $D$ l'ensemble des dimensions, $H$ l'ensemble des hiérarchies, $M$ l'ensemble des mesures, $A$ l'ensemble des attributs de dimension, $P$ l'ensemble des paramètres, $W$ l'ensemble des attributs faibles. Ces différents composants permettent la spécification d'un schéma en constellation.

**Définition 1**. Une constellation personnalisée *CP* est définie par ($F$; $D$; $Star^{CP}$; $Rule^{CP}$; $Annotate^{CP}$) où

- $F = \{F_1,…, F_n\}$ est l'ensemble des faits,
- $D = \{D_1,…, D_m\}$ est l'ensemble des dimensions,
- $Star^{CP} : F \rightarrow 2^D$ associe chaque fait à un sous-ensemble des dimensions en fonction desquelles il est analysable,
- $Preference^{CP} = \{P^{CP}_1, P^{CP}_2,…\}$ est l'ensemble de préférences de personnalisation,
- $Annotate^{CP} = \{AD^{CP}_1, AD^{CP}_2,…\}$ est l'ensemble des annotations décisionnelles.

La constellation constitue une généralisation de la modélisation en étoile (Kimball, 1996) ; si $|F| = 1$ alors C est une étoile. Cette définition décrit de manière classique la constellation au travers des concepts de fait, de dimension et de hiérarchie. L'extension réside dans la capacité d'une constellation de matérialiser l'expertise des décideurs par le mécanisme d'annotation ($Annotate^{CP}$) et par l'intégration de préférences ($Preference^{CP}$) permettant sa personnalisation.

**Exemple**. La figure 1 présente un exemple de constellation avec les formalismes graphiques associés aux concepts de fait $F = \{FVENTES\}$ et de dimension $D = \{DCLIENTS, DPRODUITS, DTEMPS\}$. On notera que $|F| = 1$, il s'agit du cas particulier du schéma en étoile où la constellation est formée d'un seul fait.

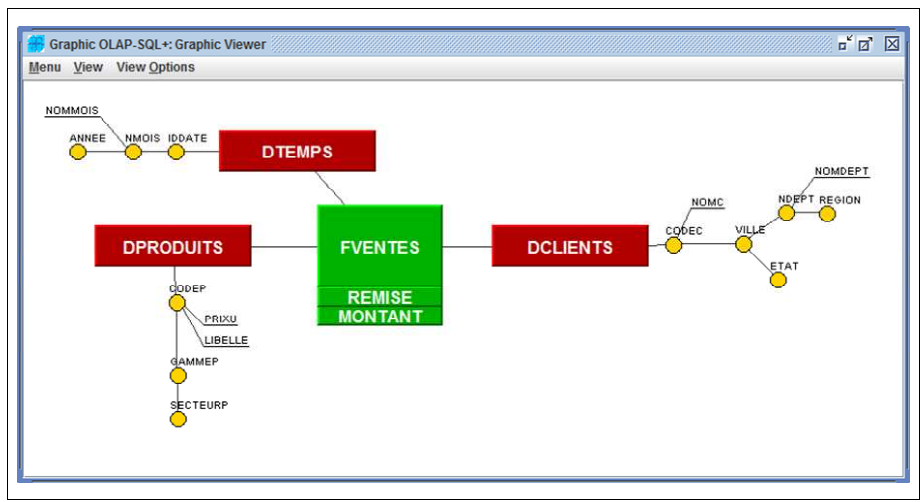

**Figure 1.** *Exemple de constellation*

**Définition 2**. $\forall i \in [1..n]$ un *fait* $F_i$ est défini par ($NF_i$ ; $M_i$) où

- $NF_i$ est le nom identifiant le fait dans la constellation,
- $M_i = \{m_1,…, m_{xi}\}$ est l'ensemble des mesures.

**Exemple**. Dans l'exemple présenté à la figure 1, le fait *FVENTES* est composé de deux mesures *REMISE* et *MONTANT*. Il est décrit de la manière suivante: (*'FVentes'* ; *{REMISE, MONTANT}*).

---

**Définition 3**. $\forall i \in [1..m]$ une *dimension* $D_i$ est définie par ($ND_i$ ; $A_i$; $H_i$) où

− $ND_i$ est le nom identifiant la dimension dans la constellation,
− $A_i = \{Id_i, All_i\} \cup P_i \cup W_i$ est l'ensemble des attributs de la dimension. On distingue les *paramètres* $P_i \subseteq P$ représentant les graduations possibles, des *attributs faibles* $W_i \subseteq W$ représentant des informations additionnelles associées aux paramètres.
− $H_i = \{H_1, ..., H_{pi}\} \subseteq H$ est l'ensemble des *hiérarchies*.

---

Les attributs de $D_i$ respectent les propriétés de disjonction $\forall (j_1, j_2) \in [1..m]^2$, si $j_1 \neq j_2$ alors $A_{j1} \neq A_{j2}$ et de recouvrement $A = \bigcup_{i=1}^{m} A_i$, $P = \bigcup_{i=1}^{m} P_i$ et $W = \bigcup_{i=1}^{m} W_i$.

---

**Définition 4**. $\forall H_j \in H_i$ une *hiérarchie* $H_j$ est définie par ($NH_j$ ; $P_{Hj}$ ; $\prec_{Hj}$ ; $Weak_{Hj}$) où

− $NH_j$ est le nom identifiant la hiérarchie dans la constellation,
− $P_{Hj} = \{p_1, ..., p_v\} \subseteq P$ est l'ensemble des paramètres de la hiérarchie,
− $\prec_{Hj}$ est une relation d'ordre sur $P_{Hj}$ telle que
  ▪ l'ordonnancement des paramètres suit un ordre total $\forall p_{k1} \in P_{Hj}, p_{k2} \in P_{Hj}, k_1 \neq k_2, p_{k1} \prec_{Hj} p_{k2} \lor p_{k2} \prec_{Hj} p_{k1}$
  ▪ il existe un paramètre *racine* $\forall p_{k1} \in P_{Hj}, Id_i \prec_{Hj} p_{k1}$
  ▪ il existe un paramètre *extrémité* $\forall p_{k1} \in P_{Hj}, p_{k1} \prec_{Hj} All_i$
− $Weak_{Hj} : P_{Hj} \rightarrow 2^{WHj}$ associe les paramètres à un ensemble d'attributs faibles.

---

Les hiérarchies respectent les propriétés de disjonction $\forall i_1 \in [1..m], \forall i_2 \in [1..m]$, si $i_1 \neq i_2$ alors $H_{i1} \neq H_{i2}$ et de recouvrement $H = \bigcup_{i=1}^{m} H_i$.

**Exemple**. Dans l'exemple présenté à la figure 1, la dimension *DCLIENTS* est décrite de la manière suivante: (*'DClients'* ; *{CODEC, NOMC, VILLE, ETAT, NDEPT, NOMDEPT, REGION}* ; *{HGEOFR, HGEOUS}*). La dimension comprend cinq paramètres *CODEC*, *VILLE*, *ETAT*, *NDEPT*, *REGION*, deux attributs faibles *NOMC*, *NOMDEPT*, et deux hiérarchies notées *HFR* et *HUS*. Ces dernières sont définies de la manière suivante :

− (*'HGEOFR'* ; *{CODEC, VILLE, NDEPT, REGION}* ; *CODEC* $\prec_{HFR}$ *VILLE* $\prec_{HFR}$ *NDEPT* $\prec_{HFR}$ *REGION* ; *{CODEC→{NOMC}, NDEPT→{NOMDEPT}}*) et
− (*'HGEOUS'* ; *{CODEC, VILLE, ETAT}* ; *CODEC* $\prec_{HUS}$ *VILLE* $\prec_{HUS}$ *ETAT* ; *{CODEC→{NOMC}}*).

### 3.2. *Analyse OLAP*

Les décideurs explorent interactivement l'espace multidimensionnel d'une constellation par une succession d'opérations de manipulation OLAP (Ravat *et al.*, 2007) afin d'obtenir le résultat souhaité. Ce résultat peut prendre la forme d'une table multidimensionnelle, d'un graphique voire la combinaison des deux. Afin de rendre notre approche indépendante des structures de visualisation, nous définissons le concept de contexte d'analyse par une description arborescente des tables multidimensionnelles (Jerbi *et al.*, 2009).

**Définition 5**. Un contexte d'analyse *CA* est défini par ($C^F$; $C^D$; $C^R$) où

– $C^F = NF_i[/f(m) \in \{[val]+\}]+$ représente le sujet (fait $F_i$) en cours d'analyse,
– $C^D = \{C^{D1}, ..., C^{Du}\}$ représente les axes de l'analyse en cours avec $\forall i \in [1..u]$, $C^{Di}$ = $ND_i[.NH_j]?[/(p_{k1}, p_{k2}) \in \{[(val_1, val_2)]+\}]+$ où $p_{k1} \in A^{Di}$, $p_{k2} \in A^{Di}$ et $(val_1, val_2) \in dom(p_{k1}) \times dom(p_{k2})$,
– $C^R = \{pred^F, pred^{D1}, ..., pred^{Du}\}$ est l'ensemble des prédicats définissant les restrictions sur les valeurs analysées.

Associé au concept de contexte d'analyse, nous définissons un formalisme sous une forme arborescente comme l'illustre la figure 2. L'arbre comporte des nœuds représentant les composants structurels (fait, mesure, dimension, hiérarchie, paramètre et attribut faible) ainsi que les valeurs. Sur les nœuds représentant les propriétés (mesure, paramètre, attribut faible), il est possible de placer des prédicats de restriction du domaine des valeurs analysées dans le contexte d'analyse modélisé.

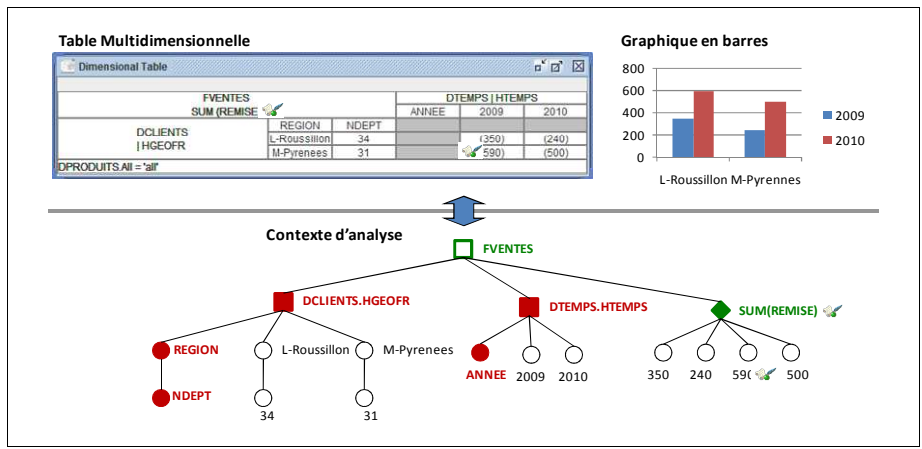

**Figure 2.** *Exemple de contexte d'analyse*

Ainsi un contexte d'analyse représente l'état courant de l'analyse menée par un usager décideur. L'analyse OLAP est vue comme une succession de contextes d'analyse sur lesquels le décideur applique des opérations de manipulation OLAP formant une séquence de contextes d'analyse où les nœuds sont les contextes d'analyse et les arcs sont les opérations de manipulation OLAP appliquées.

**Exemple**. Le contexte d'analyse présenté à la figure 2 est obtenu en appliquant la succession d'opérations de manipulation OLAP suivante. La première opération permet d'afficher une première table multidimensionnelle affichant la somme des remises en fonction de la région et de l'année du client. La seconde opération (forage) permet d'offrir un niveau de détail supplémentaire en ajoutant les informations sur le département du client.

DISPLAY(FVENTES, {SUM(REMISE)}, DCLIENTS, HGEOFR, DTEMPS, HTEMPS) = $CA_1$
DRILLDOWN($CA_1$, DCLIENTS, NDEPT) = $CA_2$

Nous ne détaillons pas dans cet article notre algèbre OLAP (*cf.* (Ravat *et al.*, 2008) (Teste, 2009)).

**4. Annoter une constellation ou une analyse**

Les prises de décisions reposent non seulement sur les données brutes mais également sur les réflexions des décideurs voire la confrontation de différentes interprétations. Notre proposition consiste à modéliser au travers d'annotations le capital immatériel mentalement associé aux données par les décideurs.

**4.1.** *Annotations*

Les annotations visent à conserver les commentaires et discussions formulés lors des analyses et du processus de prise de décisions. Ce cadre informatique permet d'exploiter et de partager les données multidimensionnelles tout en supportant des fonctionnalités d'annotation permettant d'enrichir interactivement les composants d'une constellation. Les décideurs sont alors des usagers actifs créant leur propre système de repérage au travers de signes graphiques (sur-lignage, cerclage…), de commentaires et de réponses (affirmation, infirmation) pouvant impliquer des fils de discussions (communication asynchrone).

L'expertise que véhiculent ces annotations est utilisée à des fins personnelles ou collectives et elles peuvent contribuer à améliorer les analyses futures. Les annotations contiennent des informations subjectives et des informations objectives.

> **Définition 6**. Une *annotation* $AD^{CP}_x$ est définie par ($IS^{AD}_x$; $IO^{AD}_x$) où
> - $IS^{AD}_x$ est un ensemble d'informations subjectives regroupant :
>   - le contenu textuel saisi par le décideur qui annote,
>   - le type de l'annotation (commentaire, question, réponse, conclusion).
> - $IO^{AD}_x$ est un ensemble d'informations objectives comportant :
>   - son identifiant,
>   - sa date de création permettant d'ordonner chronologiquement les annotations,
>   - l'identifiant de son créateur (décideur),
>   - référence à une annotation père,
>   - son point d'ancrage spécifiant la localisation précise de l'annotation.

Le point d'ancrage peut être de deux natures :
- un point d'ancrage *global* localisé sur un concept dans une constellation (l'annotation sera présente dans tous les contextes d'analyse intégrant le concept annoté globalement), ou bien,
- un point d'ancrage *local* localisé sur un élément dans un contexte d'analyse (l'annotation n'est présente que dans ce contexte).

> **Définition 7**. Un *point d'ancrage* $\alpha$ est défini par ($S$; $D_1$; $D_2$) où
> - $S = [CA_k.]?[ NF_i[.f(m)[=val]?]?]?$ désigne un ancrage relatif au fait $F_i$,
> - $D_1 = \lambda \mid ND_{i1} [.NH_{j1}[/pk_1[=pos_1]?]?]^*?$ désigne un ancrage relatif à une dimension,
> - $D_2 = \lambda \mid ND_{i2} [.NH_{j2}[/pk_2 [=pos_2]?]?]^*?$ désigne un ancrage relatif à une dimension.

Notons que $CA_k$ désigne un contexte d'analyse, $f(m)$ est une mesure associée à une fonction d'agrégation, val représente une valeur prise par la mesure, $pk_1$ (respectivement $pk_2$) désigne un paramètre, $pos_1$ (respectivement $pos_2$) représente une valeur prise par le paramètre $pk_1$ (respectivement $pk_2$).

La conservation de l'expertise du décideur permet d'intervenir à deux niveaux :
- Au niveau schéma. Les annotations facilitent le processus de prise de décision par une plus grande compréhension de la sémantique des composants et des instances d'une base de données multidimensionnelle.
- Au niveau analyses décisionnelles. La spécification d'analyses décisionnelles au travers d'un contexte d'analyse enrichi d'une réflexion critique s'apparente au concept de lecture active (Adler *et al.*, 1972). La tâche des décideurs est facilitée par la conservation de la réflexion critique du décideur, mais également par le partage de ces réflexions entre les différents décideurs et experts.

### 4.2. *Exploitation des annotations*

L'usage des annotations se décline selon des modalités soit personnelle, soit collective. Dans un *usage personnel*, les annotations matérialisent la réflexion et l'analyse de l'usager décideur rendant leur réutilisation possible. Dans un *usage collectif* plusieurs décideurs interviennent. Lorsque par exemple l'analyse est complexe, l'avis d'un autre expert est souvent sollicité, ce qui peut donner lieu à des débats argumentés visant à atteindre un consensus pour une prise de décision collégiale. Le support par des annotations de cet échange permet de sauvegarder et réutiliser les expertises.

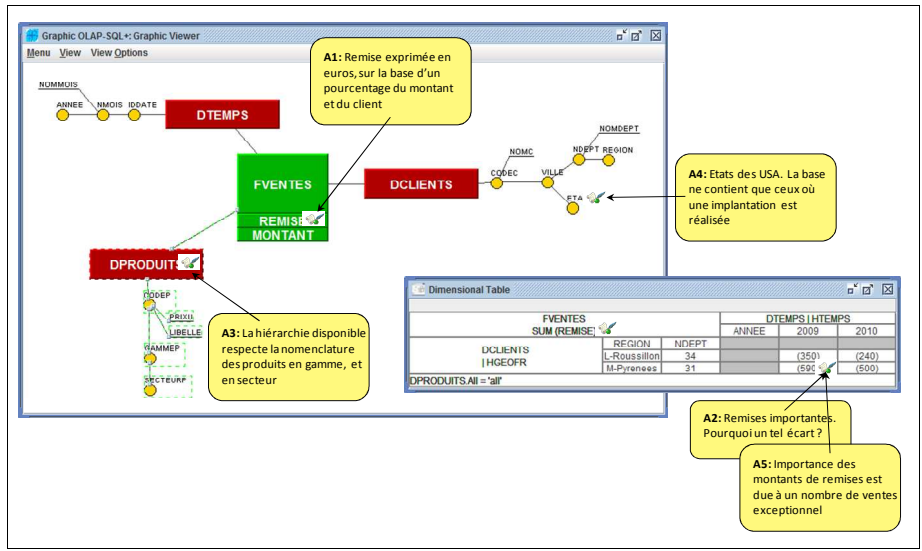

**Figure 3.** *Exemple d'annotations globales et locales*

**Exemple**. La figure 3 présente un exemple de constellation comportant des annotations. Cette constellation permet d'analyser le montant et les remises des ventes (fait *FVENTES* avec les mesures *REMISE* et *MONTANT*) en fonction du temps, du client et du produit (dimensions *DTEMPS*, *DCLIENTS* et *DPRODUITS*).

La constellation comporte trois annotations globales ($A_1$, $A_3$, $A_4$) et deux annotations locales ($A_2$, $A_5$). Suite à une première analyse, un décideur $U_1$ annote le contexte d'analyse ($A_2$) tandis qu'un second décideur $U_2$ annote ce même contexte d'analyse ($A_5$) en réponse à l'annotation $A_2$. Les ancres de ces annotations sont définies par les expressions où $CA_2$ désigne le contexte d'analyse correspondant à la table multidimensionnelle de la figure 3 :

- $A_1$ : (FVENTES/Remise, $\lambda$, $\lambda$).
- $A_2$ : ($CA_2$.FVENTES/Remise, DCLIENTS.HGEOFR/Region='M-Pyrenees', DTEMPS.HTEMPS/Annee=2009).
- $A_3$ : ($\lambda$, DPRODUITS, $\lambda$).
- $A_4$ : ($\lambda$, DCLIENTS.HGEOUS/Etat, $\lambda$).
- $A_5$ : L'ancre est identique à $A_2$, seul le contenu des annotations diffère.

## 5. Personnalisation de la constellation

En complément des annotations donnant des informations sur les données manipulées par le décideur, nous souhaitons limiter l'ensemble des données analysées aux données « préférées ». L'objet de cette section est de spécifier ce concept.

L'usager peut exprimer des préférences, notées simplement $P_i$, sur les éléments de structure d'une constellation et/ou sur les valeurs. Ces préférences peuvent être absolues ou contextuelles : une préférence absolue est toujours prise en compte par le système, tandis qu'une préférence contextuelle est prise en compte par le système lorsque le contexte d'analyse courant couvre le contexte de la préférence (Jerbi *et al.*, 2009).

---

**Définition 8**. Une *préférence* $P_i$ est définie par $(\succ_{Pi}; C^{Pi})$ où

– $\succ_{Pi}$ est une relation d'ordre sur un ensemble $E$ d'éléments.
  - Si les éléments de $E$ sont des éléments de structure de la constellation, $E \in \{F; D; H; M; A\}$, alors $P_i$ est une préférence de structures,
  - Si les éléments de $E$ sont des prédicats sur les propriétés de la constellation, alors $P_i$ est une préférence de valeurs.

– $C^{Pi} = (C^F; C^D; C^R)$ est le contexte de la préférence. Le contexte de préférence est défini comme un contexte d'analyse, pouvant comporter des parties vides.

---

**Exemple**. On considère les préférences d'un décideur : ($P_1$) « Je préfère analyser les remises par gamme, puis par code des produits », ($P_2$) « Lors de l'analyse des ventes, je préfère observer toutes les mesures simultanément pour mieux corréler les remises et les montants », et ($P_3$) « Lorsque interviennent les clients, je m'intéresse à la région Midi-Pyrénées ».

Les préférences peuvent aussi bien porter sur le schéma de la constellation que sur les valeurs de cette dernière. Par exemple, les préférences $P_1$ et $P_2$ portent sur la schéma tandis que la préférence $P_3$ concerne les valeurs. En outre, la préférence $P_1$ est une préférence absolue valide pour tous les contextes (le contexte vide est représenté par le triplet $(\lambda; \emptyset; \emptyset)$) tandis que $P_2$ et $P_3$ sont contextuelles. Les préférences sont définies par les expressions suivantes :

($P_1$) : (GAMMEP $\succ_{P1}$ CODEP; $(\lambda; \emptyset; \emptyset)$)
($P_2$) : (MONTANT $\succ_{P2}$ REMISE; (FVENTES; $\emptyset$; $\emptyset$))
($P_3$) : (DCLIENTS.REGION = ' M-Pyrenees' ; ($\lambda$ ; {DCLIENTS}; $\emptyset$))

## 6. Recommandations enrichies d'annotations

Notre objectif est d'assister le décideur dans sa navigation au sein de l'espace multidimensionnel. Ainsi à partir d'un contexte d'analyse courant $CA_i$, le décideur transforme $CA_i$ en $CA_{i+1}$ par l'application d'une opération OLAP $Op_i$. Comme l'illustre la figure 4, le système de recommandation OLAP cherche à déterminer un ensemble de recommandations $\mathcal{R}_i$ :
- par anticipation en déterminant un futur contexte d'analyse $CA_{i+j}$ que l'usager est susceptible de construire, $\mathcal{R}_i=\{CA_{i+j}\}$,
- par alternatives en suggérant les contextes d'analyse $CA_k$ alternatifs à la navigation de l'usager.

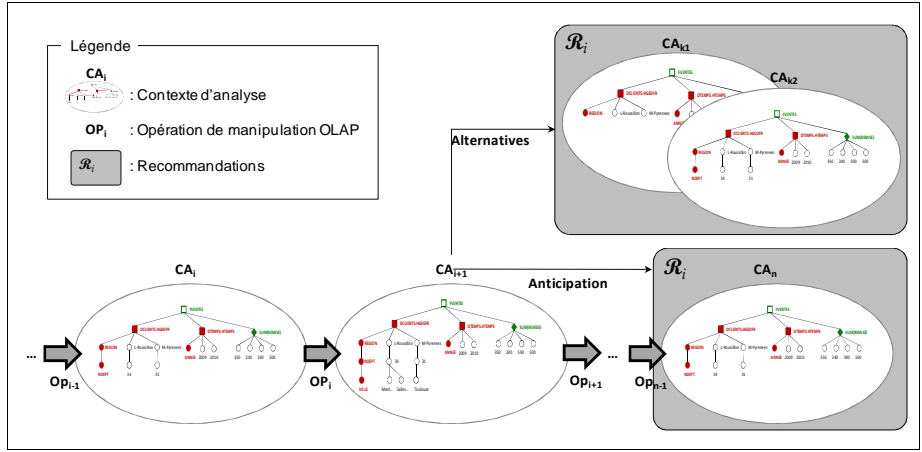

**Figure 4.** *Principe des recommandations de contextes d'analyse*

Nous nous limitons dans la suite aux recommandations alternatives. Elles consistent à proposer des contextes d'analyse considérés comme nouveaux dans l'analyse de l'usager. Pour ce faire, le contexte d'analyse $CA_{i+1}$ qui est calculé sert à déterminer un ensemble $\mathcal{R}_i$ de contextes d'analyse à recommander sur la base de préférences candidates $\mathcal{P}_i$. A partir des préférences candidates $P_j \in \mathcal{P}_i$, une recommandation $CA_k$ est déterminée en intégrant la préférence dans $CA_{i+1}$ formant un ensemble de recommandations $\mathcal{R}_i$.

```
Algorithme de recommandation d'alternatives.
Entrées : CAᵢ - Opᵢ
Sortie :  𝓡ᵢ

Début
    CAᵢ₊₁←ConstruireContexte(CAᵢ; Opᵢ);
    𝓟ᵢ←PreferencesCandidates(CAᵢ₊₁; PreferenceCP);
    𝓡ᵢ←∅;
    Pour Chaque Pⱼ∈ 𝓟ᵢ Faire
        CAₖ←CAᵢ₊₁⊕Pⱼ;
        𝓡ᵢ←𝓡ᵢ∪{CAₖ};
    FinPour;
Fin.
```

**PreferencesCandidates(CA$_{i+1}$; Preference$^{CP}$).** Le système construit l'ensemble des préférences candidates $\mathcal{P}_i$ en sélectionnant dans l'ensemble des préférences $Preference^{CP}$, celles dont le contexte est couvert par le contexte d'analyse $CA_{i+1}$. La couverture $\chi^{Pj}$ d'une préférence $P_j$ correspond à la couverture entre l'arbre du contexte d'analyse et celui du contexte de la préférence. Nous ne considérons dans cet article que les cas de couverture totale : un contexte de préférence est couvert par le contexte d'analyse $CA_{i+1}$ si et seulement si tous les arcs $(v_{k1}, v_{k2})$ appartiennent à $CA_{i+1}$. L'affichage d'un contexte d'analyse recommandé est couplé avec les annotations correspondantes. Le système extrait les annotations qui se rapportent au contexte d'analyse entier, ou à des parties de ce contexte. Ces annotations sont restituées selon la structure de visualisation utilisée.

**Exemple**. Considérons le contexte d'analyse $CA_2$. Le décideur modifie ce contexte d'analyse en changeant l'axe d'analyse des clients par celui des produits par l'opération ROTATE($CA_2$ ; DCLIENTS ; DPROUITS ; HPROD) = $CA_3$. Le contexte d'analyse obtenu permet l'analyse des remises en fonction des produits et du temps.

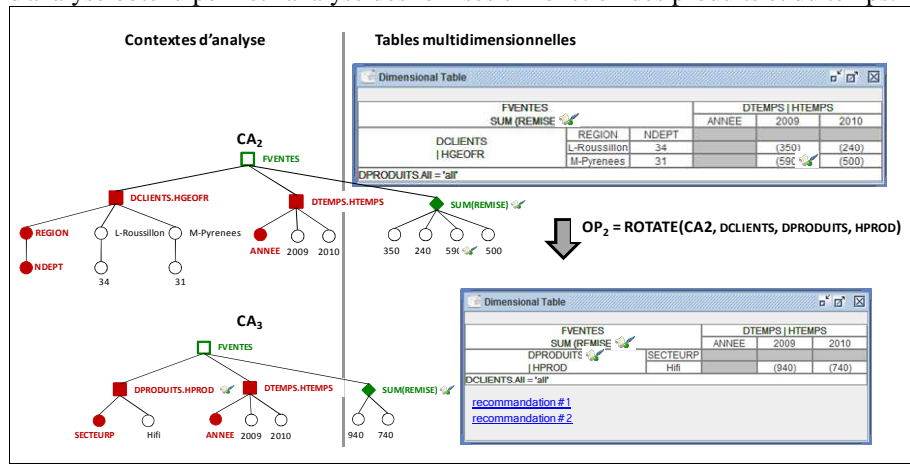

**Figure 5.** *Exemple de navigation OLAP avec recommandations annotées*

Comme l'illustre la figure 5, la table multidimensionnelle équivalente au nouveau contexte d'analyse $CA_3$ est enrichie par deux recommandations alternatives. Ces recommandations ont été calculées à partir des préférences $Preference^{CP} = \{P_1 ; P_2 ; P_3\}$ disponibles dans la constellation. Le système OLAP détermine les préférences candidates $\mathcal{P}_i = \{P_1 ; P_2\}$. Les préférences $P_1$ et $P_2$ sont candidates puisque leurs contextes de préférence respectifs ($\lambda;\varnothing;\varnothing$) et (FVENTES; $\varnothing$; $\varnothing$) couvrent totalement le contexte d'analyse $CA_3$. Par contre, la préférence $P_3$ n'est pas candidate puisque son contexte de préférence ($\lambda$ ; {DCLIENTS}; $\varnothing$) ne couvre pas le contexte d'analyse $CA_3$. Comme l'illustre la figure 6, ces préférences candidates sont utilisées pour calculer les recommandations $\mathcal{R}_i = \{CA_{k1} ; CA_{k2}\}$ en les intégrant à tour de rôle dans le contexte d'analyse courant $CA_3$.

Les annotations $A_1$ et $A_3$ sont restituées puisqu'elles sont ancrées aux contextes d'analyse recommandés respectivement la mesure Remise du fait FVENTES et la dimension DPRODUITS.

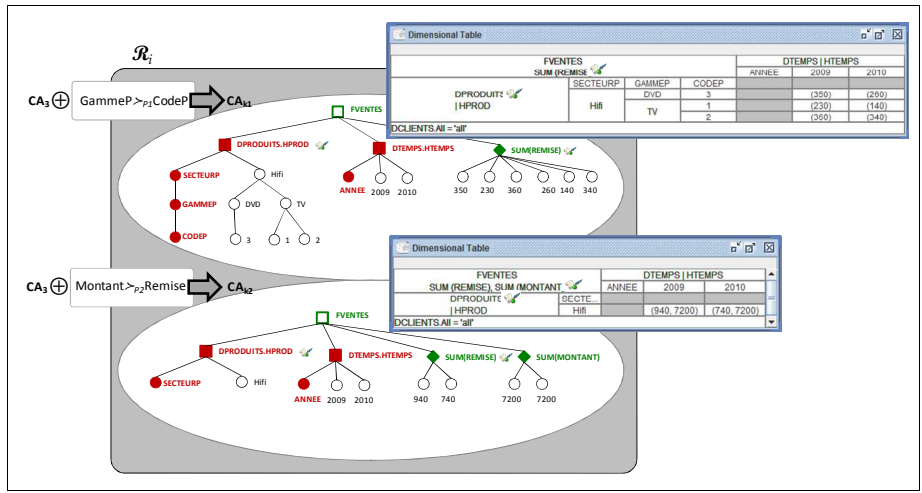

**Figure 6.** *Exemple de recommandations de contextes d'analyse alternatifs*

## 7. Conclusion

L'objectif de nos travaux sur la personnalisation dans les systèmes OLAP est double. Il s'agit dans un premier temps de rendre disponible et accessible toute information ayant permis d'aboutir à une décision. Dans un deuxième temps, nous souhaitons mieux prendre en compte les préférences de l'usager en termes de données. Ainsi, nous limiterons l'ensemble des données multidimensionnelles aux données « préférées » par les décideurs.

Une première proposition repose sur le concept de mémoire d'expertise afin de conserver le patrimoine immatériel des décideurs au sein du système OLAP. En effet, l'information utile lors du processus d'analyse décisionnelle ne se trouve pas uniquement dans les bases de données multidimensionnelles, mais une partie importante est habituellement immatérielle : il s'agit de « l'expertise » du décideur. Nous avons proposé de modéliser sous la forme d'annotations ancrées dans la base de données multidimensionnelles toutes ces informations immatérielles relevant de l'expertise de l'usager décideur (commentaires, discussions, prises de décision…).

La seconde proposition consiste en la définition d'un modèle de préférence pour mieux représenter les besoins de l'usager en matière de données analysées. Nos travaux reposent sur une approche qualitative représentant les préférences de l'usager par une relation d'ordre exprimée sur les données. Ces préférences sont alors simplement définies les unes par rapport aux autres. Nous avons représenté son contexte d'analyse pour déterminer durant l'analyse les préférences relevant de l'analyse en cours. Cette « contextualisation » des préférences permet lors des manipulations OLAP des recommandations contextuelles annotées qui assistent l'usager dans son exploration de l'espace multidimensionnel. L'assistance que nous proposons consiste à recommander à l'usager des requêtes alternatives pour obtenir des perspectives d'analyses complémentaires à son analyse.

Les extensions possibles à ces travaux peuvent se situer au niveau de l'optimisation des analyses OLAP. En effet, afin de restituer le plus rapidement possible les données multidimensionnelles, les mécanismes des vues auxiliaires d'optimisation couramment utilisés dans les systèmes OLAP doivent être adaptés afin d'intégrer les préférences annotées.

**8. Bibliographie**